\RequirePackage[T1]{fontenc}
\documentclass[conference]{IEEEtran}
\IEEEoverridecommandlockouts

\usepackage{mathptmx}
\usepackage{amsmath,amssymb}
\usepackage{booktabs}
\usepackage{graphicx}
\usepackage{xcolor}
\usepackage{tikz}
\usetikzlibrary{arrows.meta,positioning,shapes.geometric,calc}
\usepackage[draft]{hyperref}
\usepackage{url}

\definecolor{serB1}{HTML}{D55E00}  % B1 none         (vermillion)
\definecolor{serB2}{HTML}{009E73}  % B2 pre-vet      (bluish green)
\definecolor{serB3}{HTML}{E69F00}  % B3 no deadline  (orange)
\definecolor{serB4}{HTML}{0072B2}  % B4 ours         (blue)

\newif\ifdraftmarks
\draftmarksfalse                % clean submission build

\AtEndDocument{\typeout{==== TOTAL PAGES: \thepage\space ====}}
\ifdraftmarks   % where the pages actually go, so trimming is targeted
  \makeatletter
  \let\oldsection\section
  \renewcommand{\section}{\@ifstar\pgmapstar\pgmapplain}
  \newcommand{\pgmapstar}[1]{\typeout{PAGEMAP p\thepage: #1}\oldsection*{#1}}
  \newcommand{\pgmapplain}[1]{\typeout{PAGEMAP p\thepage: #1}\oldsection{#1}}
  \makeatother
\fi

\newcommand{\Eu}{E_u}
\newcommand{\Ep}{E_p}
\newcommand{\Tp}{T_p}
\newcommand{\dhide}{\delta_{\text{hide}}}

\AddToHookNext{shipout/foreground}{%
  \begin{tikzpicture}[remember picture,overlay]
    \node[anchor=south,inner sep=0pt,outer sep=0pt,text width=\textwidth,
      font=\fontsize{8pt}{9.6pt}\selectfont,align=left]
      at ([yshift=7mm]current page.south) {%
      Author accepted manuscript. Accepted for publication at IEEE CBDCom 2026.\par
      \textcopyright\ 2026 IEEE. Personal use of this material is permitted.
      Permission from IEEE must be obtained for all other uses, in any current
      or future media, including reprinting/republishing this material for
      advertising or promotional purposes, creating new collective works,
      for resale or redistribution to servers or lists, or reuse of any
      copyrighted component of this work in other works.};
  \end{tikzpicture}%
}

\begin{document}
% PDF document information (XeTeX/xdvipdfmx special; no links, no outline).

\title{Bounded Provisional Visibility: Controlling Poisoning Exposure in
Continuously Ingested RAG Vector Stores}

% Six independent author blocks, read left-to-right across the first row,
% then left-to-right across the second row, as in the supplied IEEE template.
\makeatletter
\newcommand{\linebreakand}{\end{@IEEEauthorhalign}\hfill\mbox{}\par
\mbox{}\hfill\begin{@IEEEauthorhalign}}
\makeatother
\newif\ifanon
\anonfalse
\ifanon
  \author{\IEEEauthorblockN{Anonymous Author(s)}
  \IEEEauthorblockA{Paper under double-blind review}}
\else
\author{%
\IEEEauthorblockN{Chuhong Xu}
\IEEEauthorblockA{\textit{Sony Corporate of America}\\
San Jose, USA\\
chuhong.xu@sony.com}
\and
\IEEEauthorblockN{Lu Yi}
\IEEEauthorblockA{\textit{Google LLC}\\
Mountain View, USA\\
annabelyi@google.com}
\and
\IEEEauthorblockN{Gangzhen Qian}
\IEEEauthorblockA{\textit{Google LLC}\\
Mountain View, USA\\
irisqian@google.com}
\linebreakand
\IEEEauthorblockN{Kainan Zhou}
\IEEEauthorblockA{\textit{Google LLC}\\
Mountain View, USA\\
zhoumark@google.com}
\and
\IEEEauthorblockN{Zhaoyi Li}
\IEEEauthorblockA{\textit{Independent Researcher}\\
Mountain View, USA\\
lzy9776@gmail.com}
\and
\IEEEauthorblockN{Hang Xiao}
\IEEEauthorblockA{\textit{Fortinet, Inc.}\\
Sunnyvale, USA\\
xhang@fortinet.com}}
\fi

% Author-requested local title spacing; retain the class and page geometry.
\IEEEaftertitletext{\vspace{-2.5\baselineskip}}
\maketitle

% Activates the @IEEEtranBSTCTL entry in references.bib (forced "et al.").
% Emits no visible mark; without it the control entry has no effect.
\bstctlcite{BSTcontrol}

\begin{abstract}
Continuous ingestion can expose new retrieval-augmented generation (RAG)
content to retrieval before vetting
completes, creating a temporal attack surface that conventional admission
decisions do not capture. We present a fail-closed
provisional-visibility protocol that admits new content under a deadline,
hides items whose verification has not committed in time, and supports
lineage-scoped containment. The protocol bounds unvetted exposure by the
configured visibility budget plus query-path enforcement delay;
poisoning exposure remains conditional on verifier correctness.
We evaluate the design on five workloads of encoded natural-language documents
using an exact backend and Milvus. Under verifier backlog, the deadline reduced
median poisoned retrievals from 34 (29--34) to 7 (6--7) relative to
asynchronous admission without a deadline, with the same reduction in
displaced clean results. Verify-before-visible avoided provisional exposure
but delayed clean first visibility to 6.3\,s, whereas provisional admission
made content visible within 3\,ms; expired clean items could nevertheless
experience temporary availability gaps. Replaying decisions from a
recipe-specific detector illustrated the protocol boundary: false promotions
left poison visible, while false refusals excluded clean content from retrieval.
Standalone
Milvus tests characterized enforcement delay and concurrent HNSW retrieval on
one node. These results clarify which guarantees come from the protocol and
which outcomes depend on detector quality, while quantifying the
freshness and availability trade-offs.
\end{abstract}

% One blank text line separates Abstract from Index Terms.
\par\addvspace{\baselineskip}

% Order chosen so that the block sets in two full lines under the kit's class
% with no hyphenated keyword and no one-word last line (all 720 orders were
% tested); the relative order of the original list is kept within each line.

% ==========================================================================
\section{Introduction}

Retrieval-augmented generation (RAG) grounds a language model's output on
documents
retrieved from an external knowledge base~\cite{rag}, and dense passage
retrieval turns that lookup into embedding-similarity search, commonly served
by a vector store~\cite{dpr}. As deployments move from static corpora to live
knowledge, that store must absorb new content while serving queries.

Streaming RAG refreshes the knowledge base as content
arrives~\cite{streamingrag}, while versioned vector lakes retain temporal views
over continuous updates~\cite{livevectorlake}. Both aim to reduce freshness
delay. Low freshness delay and trust assessment are in tension: establishing
that a
newly arrived item is trustworthy takes time because it requires checking
provenance, scanning for adversarial or promotional manipulation, and comparing
against existing knowledge. A freshness objective instead pushes the same item
toward the query path as soon as possible. Systems with aggressive freshness
objectives may therefore permit provisional content to become
query-visible before verification completes. This is a design option, not a
requirement: keeping unverified content invisible until fully vetted avoids
unvetted exposure at the cost of verification-dependent freshness delay, a
policy we retain as a baseline. We study the security consequences of the
low-latency operating point and how to constrain them.

Admitting content before it is vetted creates an exposure interval.
Corpus-poisoning attacks can steer dense retrievers with a small injected set,
including for unseen queries~\cite{corpuspoison}, and PoisonedRAG further
shows that a few injected texts can induce attacker-chosen answers in an
end-to-end RAG pipeline~\cite{poisonedrag}.
In a continuously ingested store, that exposure is fundamentally
temporal: a provisionally admitted poisoned vector can be retrieved
throughout the window between first visibility and containment, whether or not
a detector eventually flags it.

This temporal systems view falls between existing lines: poisoning defenses
report detection effectiveness on static or batch-constructed knowledge bases,
and continuously updated vector stores are mature but security-agnostic
(Sec.~\ref{sec:related}).

We present a fail-closed provisional-visibility protocol for streaming
vector stores. New content is query-visible but bounded by a configurable
visibility deadline $\Tp$; only positive multi-signal verification
promotes an item; expiry hides it instead. A deadline scheduler
decoupled from the verification queue makes that guarantee survive verifier
backlog, and content later found malicious is excluded from retrieval by
lineage-scoped containment without rebuilding the store. Sweeping $\Tp$ exposes
a measurable trade-off between freshness and exposure.

This paper makes four contributions. C1 (Formalization). We define
unvetted exposure $\Eu$, poisoning exposure $\Ep$, containment,
and recovery as distinct intervals, separating what a protocol can guarantee
($\Eu$) from what depends on detector quality ($\Ep$).
C2 (Protocol). We design a fail-closed provisional-visibility
protocol whose invariants I1--I7 guarantee that unverified content cannot
remain query-visible beyond $\Tp + \dhide$ and that each item is provisionally
visible at most once. A metadata time-to-live alone does not give this:
an expiry that queues behind the verifier inherits its backlog.
C3 (Containment). We implement lineage-scoped containment that
excludes a compromised item's descendants from the query path before physical
purging completes, and measure its collateral revocation.
C4 (Evaluation). We quantify deadline effectiveness and clean-content
cost, evaluate a recipe-specific feasibility detector and replay its decisions
through the promotion path, characterize eligibility completeness,
and evaluate $\dhide$ and hierarchical navigable small world (HNSW) behavior on
standalone Milvus.

% ==========================================================================
\section{Background and Threat Model}

\looseness=-1
A RAG pipeline embeds a query, retrieves the top-$k$ nearest chunks from a
vector store and conditions generation on them; under continuous ingestion an
upstream stream is chunked, embedded and inserted while queries are served. We
model the store as an insert path and a query path that observe a per-item
visibility state.

\subsection{Threat model}
\label{sec:threat}
We assume an adversary who can submit content into the ingestion stream or
control an already-authorized source. The adversary may optimize against the
retriever and any known lightweight admission signal. Prior work shows high
attack success from a few retrieval-targeted texts~\cite{poisonedrag};
SilentRetrieval further demonstrates a fluency-constrained evasive
variant~\cite{silentretrieval}. Injection is at a bounded rate
(0.1--5\% of the stream), steady or bursty, with retrieval results observable.
A poisoned item succeeds if it is retrieved for a target query before it
is contained.

The adversary cannot modify admission metadata, the deadline
scheduler, per-item visibility state, or provenance records; fully compromised
verifiers, a malicious control-plane operator, and embedding-model backdoors are
out of scope. We do admit trusted-source compromise: an item may
carry valid provenance yet be malicious. In RAGShield's numerical-claim
setting, an insider with valid signing credentials reduces provenance-only
admission to the no-defense attack-success rate~\cite{ragshield2}. We
therefore do not rely on provenance alone for promotion
(Sec.~\ref{sec:verify}).

\subsection{Scope}
We defend the ingestion-to-retrieval lifecycle of a single vector-store
deployment as observed through its supported query path; the detector/verifier
is a pluggable component throughout. Strict multi-replica/cache
consistency and fully distributed rollback are out of scope
(Sec.~\ref{sec:limits}).

% ==========================================================================
\section{Problem Formalization}
\label{sec:formal}

For an ingested item $x$ we distinguish the lifecycle timestamps
$t_{\text{arrive}}$, $t_{\text{visible}}$ (first retrievable through the
supported query path), $t_{\text{detect}}$, $t_{\text{contain}}$ (no longer
returned by the query path), $t_{\text{purge}}$ (physically deleted from the
live store) and $t_{\text{audit}}$. The visibility deadline is anchored to first visibility,
$t_{\text{deadline}} = t_{\text{visible}} + \Tp$; anchoring to arrival instead
shifts $\Eu$ by the small admission-to-visibility gap. \emph{Freshness delay}
is $D_f = t_{\text{visible}} - t_{\text{arrive}}$ and \emph{recovery} is
$W_{\text{recovery}} = t_{\text{purge}} - t_{\text{detect}}$; physical
live-store purge is treated separately from logical containment.

We define \emph{unvetted exposure} as the interval during which unverified
content is query-visible, with both endpoints on the query path:
$\Eu(x) = t_{\text{qp-exit}}(x) - t_{\text{visible}}(x)$, where
$t_{\text{qp-exit}}$ is when the path stops returning $x$ as unvetted, after
promotion, rejection or expiry.
We define \emph{poisoning exposure} as the interval during which a malicious
item can influence retrieval, $\Ep(x) = t_{\text{contain}}(x) -
t_{\text{first-poison-visible}}(x)$; it depends on detector false negatives and
promotion policy, so the protocol bounds it only when the verifier never
falsely promotes.
We define \emph{enforcement delay} to record that a metadata state change is not
containment until the query path
honors it: with $t_{\text{expire}}$ the deadline instant and
$t_{\text{qp-hide}}$ the moment the query path stops returning $x$,
$\dhide(x) = t_{\text{qp-hide}}(x) - t_{\text{expire}}(x)$.

The guarantee has three layers. (i) Operational bound. Expiry fires
at $t_{\text{visible}}+\Tp$ whatever the verifier is doing, because the
scheduler does not share its queue, and the query path stops returning the item
$\dhide$ later:
\begin{equation}
\Eu(x) \le \Tp(x) + \dhide(x).
\label{eq:bound}
\end{equation}
Only query-path visibility counts, so exposure ends at $t_{\text{qp-hide}}$
rather than at the earlier control-plane transition; anchoring there would put
$\dhide$ outside the interval and make~\eqref{eq:bound} true by arithmetic.
Equation~\eqref{eq:bound} separates the configured deadline from a
deployment-specific enforcement slack. The latter comprises scheduler firing,
state commit, and query-path propagation (Sec.~\ref{sec:instr}); each is finite
only while the control plane is live, so operators must provision and monitor
$\dhide$. The protocol makes that term independent of verifier backlog: a
separate scheduler and an $O(1)$ flag write provide it, although
shared-resource contention can still affect its magnitude.
(ii) Conditional poison bound. For a malicious item never falsely
promoted, the same bound applies to $\Ep$. (iii) General case. After
false promotion, exposure is governed by detector error and later containment.
We therefore retain an oracle arm as a mechanism upper bound and separately
evaluate one recipe-specific detector and its lifecycle consequences
(Sec.~\ref{sec:results}).

% ==========================================================================
\section{System Design}

\subsection{Architecture}
\label{sec:arch}

An admission control plane assigns new content a risk score $r(x)$ and
visibility budget $\Tp(x)$, then inserts it as PROVISIONAL. Verifier workers
and the deadline scheduler consume that state independently. The query path
reads an authoritative flag, admits PROVISIONAL and TRUSTED items to the same
top-$k$ competition without trust-based reranking, and excludes all other
states. Physical deletion remains off the critical path.

\subsection{Item state machine}

Fig.~\ref{fig:states} gives the atomic compare-and-set transitions; competing
verification and deadline events are ordered by commit time
(Sec.~\ref{sec:deadline}).

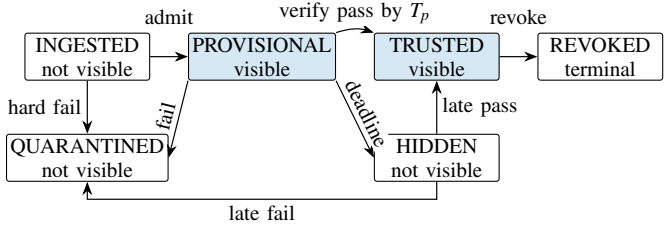
\begin{figure}[t]
\centering
\begin{tikzpicture}[
  >=Stealth,
  node distance=4.2mm and 5.0mm,
  state/.style={draw, rounded corners=1pt, minimum width=1.65cm,
    minimum height=6.2mm, align=center, inner sep=1.5pt, font=\footnotesize},
  visible/.style={state, fill=serB4!16},
  hidden/.style={state, fill=white},
  flow/.style={->, line width=.45pt},
  lab/.style={font=\footnotesize, fill=white, inner sep=1pt}
]
\node[hidden] (ing) {INGESTED\\not visible};
\node[visible, right=of ing] (prov) {PROVISIONAL\\visible};
\node[visible, right=of prov] (trust) {TRUSTED\\visible};
\node[hidden, right=of trust] (rev) {REVOKED\\terminal};
\node[hidden, below=7mm of ing] (quar) {QUARANTINED\\not visible};
\node[hidden, below=7mm of trust] (hide) {HIDDEN\\not visible};

\draw[flow] (ing) -- node[lab, above=11pt] {admit} (prov);
\draw[flow] (prov.north east) to[bend left=18]
  node[lab, above=1pt] {verify pass by $\Tp$} (trust.north west);
\draw[flow] (trust) -- node[lab, above=11pt] {revoke} (rev);
\draw[flow] (ing) -- node[lab, left=1pt] {hard fail} (quar);
\draw[flow] (prov.south west) -- node[lab, sloped, above=1pt] {fail} (quar.east);
\draw[flow] (prov.south east) -- node[lab, sloped, above=1pt] {deadline} (hide.west);
\draw[flow] (hide) -- node[lab, right=1pt] {late pass} (trust);
\draw[flow] (hide.south) -- ++(0,-2.3mm) -|
  node[lab, near start, below=1pt] {late fail} (quar.south);
\end{tikzpicture}
\caption{Protocol state machine. Shaded states are query-visible. Deadline
expiry only hides; a late pass enters TRUSTED directly, so each item has at
most one provisional episode.}
\label{fig:states}
\end{figure}

\subsection{Positive verification and hard gates}
\label{sec:verify}

Promotion requires affirmative decisions from noncompensating
source/integrity ($S$) and content/behavior ($C$) families; $C$ is mandatory
under trusted-source compromise. D1 instantiates this contract. The $S$ family checks
credential validity, provenance consistency and the pinned encoder revision.
The $C$ family sums calibration-standardized half-passage cosine repetition and distance
to the ten nearest corpus posts after excluding the item's source group. Its
threshold 1.3553 maximized poison-class F1 on 448 calibration groups, then
remained fixed for a single evaluation on 512 disjoint test groups; D1 received
neither test labels nor evaluation queries. Only the content family
contributes detector-quality evidence; the source-family controls validate
interface plumbing.

\subsection{Deadline enforcement and atomic transitions}
\label{sec:deadline}

Fail-closed enforcement rests on decoupling. A dedicated scheduler, driven by
per-item deadlines and independent of the verifier queue, performs expiration;
hiding is an $O(1)$ visibility-state transition honored by the query path
through the authoritative flag, waiting neither for verification nor for
physical deletion. The enforcement delay $\dhide$ therefore does not
queue behind the verifier, the property on which~\eqref{eq:bound} rests under
load.

Because a deadline and a verification result may occur concurrently, all
transitions are atomic compare-and-set operations conditioned on the current
state, and the race is resolved by commit time: a promotion
succeeds only if the item is still PROVISIONAL and the positive verification
committed no later than the deadline; otherwise the scheduler commits
PROVISIONAL $\rightarrow$ HIDDEN and a late positive result commits HIDDEN
$\rightarrow$ TRUSTED idempotently. A verification that finishes
computing before the deadline but commits after it is thus
post-deadline, so a verifier cannot use its internal completion time to extend
visibility.

\subsection{Lineage-scoped containment and recovery}

Each vector carries lineage metadata: document ID, source and version,
ingestion batch, embedding-model version, chunk parent, and vector ID. A
trusted item is revoked following a later verifier alert or a source/lineage
revocation event, such as a credential judged compromised or a batch flagged
by an operator. Continuous per-item large language model (LLM) monitoring is out of scope. Containment
revokes only the affected lineage rather than rebuilding the store.
Operationally, operators should choose the narrowest scope the
alert evidence justifies: source scope for a compromised publisher or
credential, batch or version scope for a suspect ingestion unit, and item scope
only when the affected object is localized and lineage is reliable. A scope
narrower than the evidence may leave correlated descendants visible, while a
broader one increases clean collateral removal.
\emph{Logical containment}, an $O(1)$ exclusion, occurs when the supported
query path no longer returns the item and ends measured poisoning exposure
($t_{\text{contain}}$); \emph{physical live-store purge} is defined separately
through recovery latency ($W_{\text{recovery}}$), so a security metric is not
conflated with database garbage collection. The query-path guarantee
establishes exclusion rather than erasure. Operationally, before live-store
purge completes, a hidden vector may still occupy index storage; snapshots or
exports can preserve it afterward. Revocation tombstones must therefore be
retained and honored during restore and rebuild for as long as such copies
exist. Access control and audit remain necessary until all retained copies are
deleted.

\subsection{Protocol invariants}
\label{sec:invariants}

Each invariant maps to a tested property. I1--I2 require affirmative $S$ and
$C$ decisions with no hard-gate failure; the oracle tests the promotion
outcome and D1 the concrete decision path. I3 makes expiry hide, I4 enforces~\eqref{eq:bound}, I5 ends exposure at logical containment, I6 decouples expiry
from verification and deletion, and I7 permits one provisional episode.
Lifecycle invariants I3--I7 run in both arms. A risk-dependent
$\Tp(x)=f(r(x))$ changes none of them.

% ==========================================================================
\section{Implementation}
\label{sec:impl}

We implement the control plane on a streaming vector-store stack. Content
arrives through an asynchronous ingestion producer and is routed by an
admission service that applies the cheap hard gate and assigns the
visibility deadline, then inserts admitted items into a single
Milvus~\cite{milvus} collection. PROVISIONAL and TRUSTED are eligibility states
in the control store rather than separate collections: promotion flips state
and moves no vector, and the query service serves both under one top-$k$
(Sec.~\ref{sec:arch}).

The authoritative item metadata
(visibility state, deadline, lineage, and audit timestamps) lives in a separate
lightweight control store, not in the vector database, which holds
vectors and serves retrieval while the query service decides eligibility.
Manu retains row-level ACID while explicitly relaxing complex transaction
requirements for vector workloads~\cite{manu}. Our protocol needs atomic
transitions over those fields, so they remain in the control store. Under
post-filtering, updating the visibility flag is an $O(1)$ write that the query
path honors immediately and that is decoupled from deletion or compaction. The timer-ordered scheduler does not
share the verifier pool; together, these two forms of decoupling realize I6.
Transitions are compare-and-set updates
(Sec.~\ref{sec:deadline}), and lineage columns enable scoped revocation without
a rebuild.

The verifier interface has two instantiations. A deterministic oracle isolates
lifecycle timing; the detector arm replays held-out D1 decisions and measured
detector-only service times through the same promotion path. It therefore
captures observed errors and latency but excludes online detector and model
contention. A single-process in-memory harness validates the control plane
without noise. The same interface can be backed by a real vector store.
Section~\ref{sec:elig} compares the resulting post-retrieval eligibility check
with an in-index filter. The harness,
preregistrations and authoritative result files are available from the
authors, each recording the runtime-code fingerprint it was produced under.

% ==========================================================================
\section{Evaluation Setup}

We ask whether $\Eu$ is bounded and how much the enforcement delay $\dhide$
widens it; whether that bound survives verifier overload; how freshness trades
against exposure over $\Tp$; what top-$k$ completeness cost eligibility
enforcement imposes; and how precise containment is. A separate experiment
reports HNSW index fidelity~\cite{hnsw} and descriptive single-node throughput
and tail latency
across five independently restarted Milvus Standalone processes
(Sec.~\ref{sec:standalone}).

\subsection{Instrumentation and derived metrics}
\label{sec:instr}
Every lifecycle metric derives from one versioned event schema of per-item
timestamps, so the standalone deployment emits the same fields without changing
a definition. Three distinctions matter. (i)
$t_{\text{verify-end}} \ne t_{\text{verify-commit}}$: a result committed after
the deadline is post-deadline (Sec.~\ref{sec:deadline}). (ii) Scheduler firing,
state commit and query-path propagation are components of $\dhide$, not extra
terms in~\eqref{eq:bound}; observed through probes,
it is interval-censored, so we report the conservative upper bound and never
charge the probe interval to system delay. (iii) Per-query item state is logged
as observed at return time, never reconstructed from the item's later
state, which is what the residual exposure rate depends on.

\subsection{Workloads, baselines, and parameters}
Four workload classes stress different facets: ingestion under concurrent
retrieval, a poison burst, a verifier slowdown exercising I6, and
post-promotion revocation. Each poison-burst workload is generated from
scratch: an eight-topic corpus of 600 unit vectors, eight queries from the
attacked topic for crafting and eight more, disjoint, for measurement, eight
from other topics as controls, and six poison vectors perturbed toward the
crafting queries alone. Queries are issued at a fixed rate over a fixed window
with the poison injected partway in, using five fixed seeds, all retained. Four baselines
share one ingestion path and differ only in
admission policy: B1 immediate admission with no verification,
B2 synchronous verify-before-visible, B3 async visible with no
deadline, and B4 the proposed protocol. B3 isolates the deadline's own effect,
differing from B4 in nothing else. The deadline $\Tp$ is swept over points bracketing the
verifier's own latency.

\subsection{Eligibility enforcement}
\label{sec:elig}
A post-retrieval authoritative check (post-filter) and an in-index metadata
filter trade off along three axes: $\dhide$, query latency, and retrieval
completeness. Post-filtering gives a small
$\dhide$ because the control store is authoritative, but it filters after approximate nearest-neighbor (ANN) search,
so it can under-fill the top-$k$; the in-index filter avoids under-fill but its
hide observability is left to the store's read-consistency level~\cite{manu},
which we fix and report. With $\alpha$ over-fetch and pool size $M=\alpha k$, post-filtering
under-fills exactly when pool hidden occupancy $h_M(q)$ exceeds
$1-1/\alpha$, so under-fill is governed by the upper tail of $h_M$ and
not its mean. We therefore vary three hidden distributions: uniform-random,
semantic-clustered, and query-coupled. The query-coupled set is built with
knowledge of the evaluation queries, so it is a worst case by
construction; the eligible population stays $\ge k$, so exact ground truth
stays well defined.
Completeness is reported as under-fill (result count $<k$) and
Eligible Recall@$k$ against exact brute-force ground truth, plus
Conditional Recall@$k$ given a full-$k$ result, which isolates ANN
approximation error.

\subsection{Invariant validation and reporting}
Automated tests assert that timeout never promotes, that each item enters
PROVISIONAL at most once, that a post-deadline-committed pass fires
HIDDEN $\rightarrow$ TRUSTED rather than PROVISIONAL $\rightarrow$ TRUSTED, that
HIDDEN and REVOKED items are never returned, and that logical containment holds
while physical purge is pending. An overload test injects heavy verifier
backlog and confirms unvetted items still leave the visible set within
$\Tp + \dhide$.

% ==========================================================================
\section{Results}
\label{sec:results}

The exact and Milvus Lite tiers validate lifecycle mechanics and eligible
ground truth; D1 adds measured decisions, and Standalone adds HNSW with
1--16-client load.

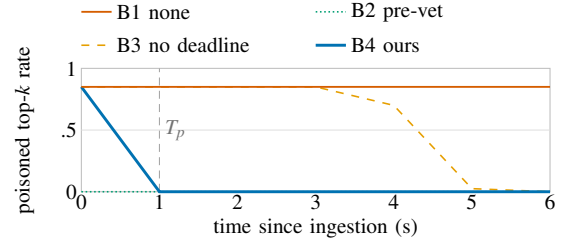
\begin{figure}[t]
\centering
%% Generated by paper/figB.py straight from prototype/results/W2-milvus.json,
%% so the figure cannot drift away from the numbers. Regenerate after any
%% re-run of poison_exposure.py; do not hand-edit figB.tex.
\begin{tikzpicture}[font=\fontsize{8pt}{9.6pt}\selectfont,inner sep=0pt,yscale=0.68]
  \draw[black!25] (0,0.00mm) rectangle (62.00mm,24.00mm);
  \draw[black!12] (0,12.00mm) -- (62.00mm,12.00mm);
  \node[left=1.5pt] at (0,0.00mm) {0};
  \node[left=1.5pt] at (0,12.00mm) {.5};
  \node[left=1.5pt] at (0,24.00mm) {1};
  \draw[black!35,densely dashed,line width=.4pt] (10.33mm,0.00mm) -- (10.33mm,24.00mm);
  \node[anchor=west,black!55] at (11.03mm,12.00mm) {$T_p$};
  \draw[serB3,dashed,line width=.6pt] (0.00mm,20.40mm) -- (10.33mm,20.40mm) -- (20.67mm,20.40mm) -- (31.00mm,20.40mm) -- (41.33mm,16.80mm) -- (51.67mm,0.60mm) -- (62.00mm,0.00mm);
  \draw[serB2,densely dotted,line width=.6pt] (0.00mm,0.00mm) -- (10.33mm,0.00mm) -- (20.67mm,0.00mm) -- (31.00mm,0.00mm) -- (41.33mm,0.00mm) -- (51.67mm,0.00mm) -- (62.00mm,0.00mm);
  \draw[serB4,solid,line width=1.1pt] (0.00mm,20.40mm) -- (10.33mm,0.00mm) -- (20.67mm,0.00mm) -- (31.00mm,0.00mm) -- (41.33mm,0.00mm) -- (51.67mm,0.00mm) -- (62.00mm,0.00mm);
  \draw[serB1,solid,line width=.7pt] (0.00mm,20.40mm) -- (10.33mm,20.40mm) -- (20.67mm,20.40mm) -- (31.00mm,20.40mm) -- (41.33mm,20.40mm) -- (51.67mm,20.40mm) -- (62.00mm,20.40mm);
  \node[below=1pt] at (0.00mm,0) {0};
  \node[below=1pt] at (10.33mm,0) {1};
  \node[below=1pt] at (20.67mm,0) {2};
  \node[below=1pt] at (31.00mm,0) {3};
  \node[below=1pt] at (41.33mm,0) {4};
  \node[below=1pt] at (51.67mm,0) {5};
  \node[below=1pt] at (62.00mm,0) {6};
  \node[below=10pt] at (31.00mm,0) {time since ingestion (s)};
  \node[rotate=90,anchor=south] at (-6.2mm,12.00mm) {poisoned top-$k$ rate};
  \draw[serB1,solid,line width=.7pt] (0.00mm,35.00mm) -- (4.00mm,35.00mm);
  \node[anchor=west] at (4.60mm,35.00mm) {B1 none};
  \draw[serB2,densely dotted,line width=.6pt] (31.00mm,35.00mm) -- (35.00mm,35.00mm);
  \node[anchor=west] at (35.60mm,35.00mm) {B2 pre-vet};
  \draw[serB3,dashed,line width=.6pt] (0.00mm,28.50mm) -- (4.00mm,28.50mm);
  \node[anchor=west] at (4.60mm,28.50mm) {B3 no deadline};
  \draw[serB4,solid,line width=1.1pt] (31.00mm,28.50mm) -- (35.00mm,28.50mm);
  \node[anchor=west] at (35.60mm,28.50mm) {B4 ours};
\end{tikzpicture}
\caption{Poisoned top-$k$ rate over time on the crafted query set (Milvus
Lite), pooled over five workloads; one bin is one query rotation, below which
the rate is undersampled by construction. B3 holds at the undefended
baseline's rate until verification drains, B4 is cut at $\Tp$, and B1 is never
contained within the window. Only the backlogged condition is drawn: with the
verifier keeping up, B3 and B4 coincide.}
\label{fig:poison}
\end{figure}

\subsection{Deadline enforcement under verification backlog}
The harness validates invariants I1--I7 and the commit-time rule (20/20
checks). The real-time functional slice runs the full
ingest $\rightarrow$ provisional $\rightarrow$ deadline $\rightarrow$ hide
$\rightarrow$ late-verify $\rightarrow$ trusted $\rightarrow$ revoke lifecycle
with 8/8 checks in both eligibility modes. Decoupling the scheduler from
verification is what bounds unvetted exposure under backlog: under $50\times$
backlog, decoupled scheduling gave $\Eu \approx \Tp$ versus $50\,\Tp$ under
coupled scheduling. On the Milvus Lite real-time control plane, $\Eu$ was
$1.00\,\Tp$ decoupled versus ${\approx}5\,\Tp$ coupled under a ${\sim}5$\,s
backlog.

\subsection{Retrieval-layer poisoning exposure and clean-content cost}
\label{sec:poison}
Poison crafted against a designated query set is injected into a collection
held loaded throughout, with eligibility on the post-filter path, and we
measure what each baseline lets it do (Fig.~\ref{fig:poison}): five
independently generated workloads whose synthetic corpora keep exact
eligible-neighbor ground truth computable;
Sec.~\ref{sec:realtext} repeats the grid on encoded natural-language documents.
Three query sets separate what one poisoned-retrieval rate conflates.
The set $Q_{\text{craft}}$, the queries the poison was optimized against, is a worst
case by construction; $Q_{\text{tgt}}$ holds disjoint queries of the
same topic, so a hit there is transfer; $Q_{\text{neg}}$ is unrelated. Rates
are reported over the queries B1 compromised, a set applied unchanged to every
baseline and $\Tp$; counts are unaffected by that conditioning.

The poison reached 75--100\% of crafted queries, while transfer was partial:
it reached 12--38\% of disjoint same-topic queries, with every workload showing
some transfer, and produced no poisoned retrieval on any of the $40$ distinct
unrelated queries.
Whether a query retrieves poison is decided by the best poison it can see
against its own top-5 cutoff, so we report that pairing rather than an average
over unrelated pairs: for the median target query the best poison falls
$0.037$--$0.061$ short of the cutoff, and yet $1$--$3$ of each workload's eight
queries clear it. Predicted and retrieved agree for every one of the $80$
held-out queries across both workloads (Sec.~\ref{sec:realtext}), so the
transfer rate is a property of the geometry, not of the defense or of when a
query happened to run.

Under backlog, the deadline rather than verification ended exposure
(Table~\ref{tab:exposure}). B1 never contains the poison; its exposure window
ends only when measurement stops. B3 holds at B1's rate for $4.7$\,s until verification drains, and
B4 is cut at the deadline; the conditional rates are $0.57$--$0.63$ for B3 and
$0.125$ for B4, versus $0.875$ for B1.
Counts and rates are both per-workload medians over workloads whose
qualified-query counts differ, so the rate is not the ratio of the
counts. When the verifier keeps up, the two are indistinguishable, because
verification commits inside $\Tp$ and the deadline never binds: the deadline is
what makes the bound survive backlog, not what creates it.

\begin{table}[t]
\caption{Median poisoned retrievals on $Q_{\text{craft}}$ across five
workloads, cumulative clean-result displacement, and the $\Tp$ sweep.
Ranges are shown where outcomes varied; comparisons are within workload.}
\label{tab:exposure}
\centering
\footnotesize
\setlength{\tabcolsep}{4pt}
\renewcommand{\arraystretch}{0.9}
\begin{tabular}{@{}lrr@{}}
\toprule
 & synthetic & real text \\
\midrule
\multicolumn{3}{@{}l}{\emph{poisoned retrievals, heavy verifier backlog}}\\
\quad B1 no admission control  & 49 (42--56) & 49 (42--49) \\
\quad B2 verify-before-visible & 0           & 0 \\
\quad B3 async, no deadline    & 35 (29--37) & 34 (29--34) \\
\quad B4 this work             & 7 (6--8)    & 7 (6--7) \\
\multicolumn{3}{@{}l}{\emph{same, verifier keeping up}}\\
\quad B3 async, no deadline    & 3 & 3 \\
\quad B4 this work             & 3 & 3 \\
\midrule
clean positions lost, B1 / B3 / B4   & 56 / 37 / 8 & 49 / 34 / 7 \\
B4 at $\Tp = 0.15$, $0.3$, $1$, $2$\,s     & 2, 3, 7, 14 & 1, 3, 7, 14 \\
\midrule
crafted queries reached                & 75--100\% & 75--88\% \\
held-out same-topic transfer           & 12--38\%  & 12--38\% \\
unrelated queries reached, of 40       & 0         & 0 \\
\bottomrule
\end{tabular}
\end{table}

\looseness=-1
This reduction came with a clean-content trade-off. Verify-before-visible
delays clean first visibility until verification
completes: $0.62$\,s with the verifier keeping up, $6.3$\,s under
backlog. The deadline makes clean content provisionally visible within three
milliseconds in both conditions, but does not make availability continuous or
durable: under backlog all six clean items expired before verification in
every workload; five were readmitted by the observation horizon and one was
right-censored. The median across the five per-workload median completed-gap
durations was $5.3$\,s, with each workload contributing five completed gaps.
The appropriate policy depends on application requirements: a
security-sensitive deployment can shorten $\Tp$; provisioning verification to
commit before expiry reduces gaps under expected load; and an application that
cannot tolerate gaps after first visibility should use verify-before-visible
and accept its initial freshness delay. Lengthening $\Tp$ reduces expiry risk
only by widening the unvetted-visibility budget. The deadline does not reduce
a poison's effect while it remains visible: relative to the clean top-$k$ that
would have been returned without the poison, the poison displaced $20$\% of
clean results under B1, B3, and B4 alike, and eligible retrieval returned the
exact top-$k$ after containment. Per-hit
displacement was therefore identical across the three; only duration, and
hence the cumulative figure, differs. Exposure scales with the budget, so $\Tp$
is a calibrated dial rather than a qualitative switch. These runs exercise a
real Milvus Lite collection, insertion path, top-$k$ query, and lifecycle; the
exact backend agrees on the main-grid poisoned-retrieval counts except for one
retrieval at the shortest deadline.

\subsection{Real documents, a real encoder, and poison written as text}
\label{sec:realtext}
A planted geometry cannot establish transfer to a retriever whose geometry did
not define the attack, so we repeat the grid of Sec.~\ref{sec:poison} with the data
replaced; a validation check confirmed that all timing and protocol parameters
were unchanged. The corpus becomes $600$ posts from eight
20 Newsgroups categories spanning recreation, science, computing, politics,
religion and commerce, $75$ each, with headers, footers and quoted replies
stripped, encoded by all-MiniLM-L6-v2 at $384$ dimensions; queries are the
opening words of held-out posts, leaving $Q_{\text{craft}}$ and $Q_{\text{tgt}}$
disjoint real utterances. The poison is an English passage inspired by
PoisonedRAG's black-box query-repetition construction~\cite{poisonedrag}: our
implementation repeats the target query, appends fixed assertive filler and
tests whether the result enters the encoder's top-$k$ without manual
vector-space placement. It does not instantiate PoisonedRAG's target-answer
generation condition.

The attack survives the substitution. Judged by the same per-query top-5 cutoff
rule, the poison falls $0.034$--$0.093$ short for the median target query, and
$1$--$3$ of each workload's eight clear it anyway. Poison was retrieved for
$12$--$38$\% of these queries, with at least one transfer in every workload,
but for none of the $40$ distinct unrelated queries. Prediction and retrieval
agreed for all $40$ real and all $40$ controlled queries. Query opportunity was
fixed: each query ran eight times, seven against an unchanged candidate set.
Together, these observations support attributing transfer to the encoded
geometry rather than query opportunity.

The right-hand column of Table~\ref{tab:exposure} shows the same baseline
ordering: B3 degrades under backlog while B4 holds at the deadline, the two are
indistinguishable when the verifier keeps up, and exposure remains monotone in
the budget. Counts are not directly compared across corpora and dimensions. The
oracle supplies the promotion outcome and nothing else: it does not construct
retrieval outcomes, alter embeddings, or decide whether a poisoned item enters
the top-$k$.

We next replace the oracle with controlled and measured detector outcomes. In
the controlled sweep, B4 was configured to falsely promote $k=0, 1, 2, 3$
items: B1 stayed at $49$ median poisoned retrievals, while B4 rose from $7$ to
$9$, $13$ and $15$, with every observed provisional exposure below
$1.02$\,s. On 512 held-out source groups, D1's TP/FP/TN/FN counts were
129/65/255/63; recall was 125/128 for the matched recipe but 4/64 for the
predefined natural-cover-suffix cross-attack, while 65/320 clean items were
falsely refused. Detector-only P50, P95, and P99 latencies were 26.1, 37.7,
and 49.8\,ms, respectively.
Each 300-admission HNSW replay produced 38 false promotions and 29 false
refusals. All 38 false promotions remained visible at the window end under
B2--B4; B2 never exposed false-refused clean items, whereas B3/B4 exposed them
briefly and excluded them from retrieval by the window end. The deadline bounds
provisional exposure but cannot revoke poison after promotion.

\subsection{Eligibility enforcement and top-$k$ completeness}
Values below are means across three seeds;
each seed's P95, under-fill rate and recall are computed separately before
aggregation. At 30\% global hidden occupancy with $2\times$
over-fetch, clustered hiding raised the P95 candidate-pool hidden occupancy
from $0.47$ to $0.58$ and under-fill from
$0.7$\% to $8$\% at an unchanged median $h_M$ of $0.30$; the query-coupled
stress construction gave $26$\%. At
$1\times$ over-fetch, Eligible Recall@$k$ was $0.70$, $0.68$ and $0.38$ under
uniform, clustered and query-coupled hiding.
At $4\times$, observed under-fill was eliminated and recall reached $1.0$ in all
three. Conditional Recall@$k$ was $1.0$ wherever a full-$k$ result existed and
is undefined for the query-coupled $1\times$ cell, which produced none,
so recall loss follows from under-fill rather than selection error. The
in-index filter returned the exact eligible top-$k$ wherever the eligible
population stayed $\ge k$.
Over-fetch is not free. At $4\times$, each query requests four times as many
candidates as at $1\times$, so candidate transfer and post-filter checks increase;
the backend-specific effects on ANN search cost and latency were not measured.
Because $\alpha$ never relaxes the authoritative eligibility check, operators
should use the smallest $\alpha$ that meets their completeness target for the
observed upper tail of $h_M$, or use in-index filtering when its propagation delay
and consistency semantics are acceptable. The $4\times$ result demonstrates
completeness for these workloads, not zero deployment cost.

\subsection{Milvus Lite functional validation}
On Milvus Lite, the adapter, all four state paths and both eligibility modes
execute against a real Milvus API and reproduce the trends above. Conditional
Recall@$k$ was 1.0 because the FLAT index adds no ANN approximation.

\subsection{Lineage-scoped containment}
\label{sec:revoke}
In the evaluated revocation workload, lineage-scoped containment caught 100\%
of the poison at all three granularities, while collateral removal fell from
${\sim}83$\% (by source) to ${\sim}43$\% (by batch) to 0\% at exact
item-lineage granularity. At equal poison recall, these figures quantify the
collateral cost of each scope choice and the reduction in collateral removal
when alert evidence supports a narrower scope; they do not identify a universally
optimal granularity.
Through the post-filter path, no residual retrieval followed logical
revocation on either backend. Through the in-index path, it persisted
immediately after the scalar update on Milvus Lite (${\sim}29$--48\% of probes), so
that path propagates a hide with nonzero delay (Sec.~\ref{sec:elig}).

\subsection{Standalone system results}
\label{sec:standalone}
We first conduct a single-run propagation microbenchmark on Milvus Standalone
v2.4.15 (single node, Docker)~\cite{milvus} over a $200$-item collection of
synthetic unit
vectors and measure the in-index mode's $\dhide$ end to end. The delay $\dhide$ is
computed from deadline expiry per the logged timestamps; probing begins at
acknowledgment. Two probes run concurrently: a scalar point lookup and an exact
(FLAT) top-$k$ search. FLAT is used because under an approximate index a
vanished item cannot be distinguished from a recall miss; concurrency keeps
either probe's service time from being charged to the other. The grid is four
read-consistency levels $\times$ \{no, heavy\} backlog, $\Tp = 1$\,s, 200 hide
events per cell, with an identical control cell at the start and end of
the run; we report $\dhide$ as a conservative upper bound derived from interval censoring
(Sec.~\ref{sec:instr}). The hide is an override upsert because v2.4.15 has no
in-place scalar update, so the magnitudes are properties of that update path as
well as of the protocol.

Read consistency determined whether completed probes remained stale after the
hide committed. Under Strong and Session, no completed probe returned the item
in any cell: 0 stale reads out of 200 events per cell. Under Bounded and
Eventually, provisional items stayed retrievable across dozens of
completed probes.

In the Bounded and Eventually cells $\dhide$ was 288--332\,ms at
P50 and 380--431\,ms at P95; the 9--12\,ms censoring bracket means that the
reported bounds overstate the true delay by at most ${\sim}12$\,ms. Under Strong and
Session, the probe itself costs ${\sim}420$\,ms at P95, so the first completed
probe already reports the item hidden and $\dhide$ is bounded but never
point-estimated. That 420\,ms characterizes the instrument, not the
underlying delay, which may be identical across modes. Adding the scalar probe
added no material cost to the top-$k$ search path under this workload (P95
421.0\,ms with it versus 419.7\,ms without), so these delays are not an
instrumentation artifact.

Thus $\dhide$ must be budgeted: at $\Tp = 1$\,s, a P95 enforcement delay of
${\sim}0.43$\,s widens the worst-case unvetted-exposure window by more than
40\%. That magnitude belongs to the in-index scalar-update path, not to the
control-store post-filter configuration of Sec.~\ref{sec:poison}.

The panel cannot identify a verifier-backlog effect. Its two control cells
perform identical query-only work at the run's start and end, yet their P50
values differ by 22.3\,ms (6.5\%), a shift that exceeds the between-cell effects. We
therefore report rounded magnitudes and categorical behavior only: where
$\dhide$ resolved, the backlog contrast changed sign (Bounded $+50$\,ms,
Eventually $-23$\,ms), and both magnitudes were of the same order as the control
drift. With 200 events per cell, the panel supports median estimation; P95 is
tail-limited, and P99 is not reported.

We report five Milvus process runs that passed fixed validity gates; the
artifact retains all attempts. Each run built a 100,000-vector HNSW index using
cosine similarity~\cite{hnsw}, with $M=16$,
\texttt{efConstruction}=200, and Strong consistency. Each run used the same
fixed set of 192 queries. At the primary \texttt{ef}=64 setting, run-level
median Recall@5 was 0.9979 (range 0.9979--0.9990), median P95 latency was
0.412\,s, and no query errors occurred. With 20 admissions/s, median throughput
across the four admission policies increased from 3.27--3.33 queries per second (QPS) at concurrency
1 to 48.0--51.27 QPS at concurrency 16, with no admission failures. Median P95
query latency at concurrency 16 was 0.408--0.416\,s.

% ==========================================================================
\section{Related Work}
\label{sec:related}

Corpus-poisoning attacks can steer dense retrieval with a small injected
set~\cite{corpuspoison}, and PoisonedRAG extends the threat to attacker-chosen
generation~\cite{poisonedrag}. We propose no new attack.

SilentRetrieval studies fluent, semantics-preserving variants of such
poisoning~\cite{silentretrieval}. Our threat model admits this adversary
(Sec.~\ref{sec:threat}).

RAGuard filters suspicious retrieved chunks~\cite{raguard}, and PRA-RAG
robustly aggregates retrieved evidence~\cite{prarag}.

Orchestration and forensics complement such filtering. ADO selects defenses
per query context~\cite{ado}, and RAGForensics traces harmful outputs to
responsible passages~\cite{ragforensics}. These studies predominantly use
static or batch-constructed corpora and report method-specific outcomes rather
than temporal exposure, containment and recovery under continuous ingestion.

In-toto formalizes software-supply-chain attestation for how an artifact was
produced~\cite{intoto}, and RAGShield applies provenance checks to RAG
ingestion and shows their insufficiency against a valid-key
insider~\cite{ragshield2}. The latter result motivates our multi-signal rule
without prescribing a detector.

Milvus provides similarity search over mutable collections~\cite{milvus},
and Manu develops a cloud-native vector-DBMS architecture~\cite{manu}. Our control store holds only eligibility state beside such a
system (Sec.~\ref{sec:impl}).

FreshDiskANN supports incremental updates of a graph index~\cite{freshdiskann},
while SPFresh performs in-place updates without full rebuilds~\cite{spfresh}.

Freshness has also been pursued at the RAG layer. Streaming RAG incrementally
refreshes a compact prototype index~\cite{streamingrag}, and SPI adapts
retrieval resolution per query~\cite{spi}.

Temporal retrieval is a further direction. LiveVectorLake adds versioned
point-in-time retrieval~\cite{livevectorlake}. These systems optimize
freshness, update efficiency or temporal retrieval but do not model adversarial
exposure windows.

A recent RAG-security survey separates admission-time integrity, retrieval-time
hardening and post-retrieval mitigation~\cite{ragsurvey}. Our focus is the
temporal ingestion-to-retrieval lifecycle under concurrent updates.

% ==========================================================================
\section{Limitations}
\label{sec:limits}

The $\Eu \le \Tp+\dhide$ bound inherits the deployment's enforcement delay.
The corresponding $\Ep$ bound holds only absent false promotion; falsely
promoted items require later containment. Verify-before-visible avoids
provisional exposure at higher freshness delay, and a deadline limits exposure
duration rather than damage per retrieval. Accordingly, the protocol controls
unvetted-visibility duration; it does not solve poisoning detection. Hard gates
assume an honest control plane; post-filtering requires over-fetch; distributed
clocks and rollback remain out of scope.

D1 is one recipe-specific detector evaluated on 512 held-out source groups.
Its matched-recipe and cross-attack results do not establish detector
generality, unknown-attack robustness or second-detector replication, and
offline replay excludes inference contention. The real-text study ends at
retrieval rather than answer generation. Physical-purge latency, storage
overhead while hidden vectors remain in the index, and restore or reindex
behavior were not evaluated.

The Standalone propagation panel is a single-run measurement of the in-index
update path; a second-host replication passed its fixed stability gate in only
two of five runs and therefore contributes no latency estimate. The five HNSW
runs are single-node measurements, and 98,248 of 100,000 vectors are
deterministic pressure data. Requests are not independent statistical units;
QPS and latency tails do not establish production capacity, availability,
fault tolerance, or distributed scaling. Milvus Lite and cross-backend results
validate implementation consistency, not independent performance replication.

% ==========================================================================
% Keep the short conclusion intact on the final page.
\newpage
\section{Conclusion}

\looseness=-1
Continuous RAG ingestion creates distinct unvetted- and poisoning-exposure
intervals. The protocol bounds the first by a deadline plus query-path
enforcement delay and supports lineage-scoped containment; on encoded
natural-language workloads under backlog it cut poisoned retrievals and
displaced clean results by roughly a factor of five, while clean availability could
lapse. D1's false promotions and refusals show why the poison bound is
conditional and why detector quality has a clean-content cost. Five single-node
HNSW runs show high-recall ANN retrieval under concurrent ingestion for this
workload.

% ==========================================================================
\section*{Acknowledgment}

Codex(OpenAI) was used for language editing, experiment running, and LaTeX typesetting. The authors reviewed the manuscripts, verified the reported results and citations, and take responsibility for the final text.

% ==========================================================================
% References follow the acknowledgment using the template's centered column
% heading. The author approved this layout to preserve the eight-page limit.
\IEEEtriggercmd{\newpage}
\IEEEtriggeratref{7}
\bibliographystyle{IEEEtran}
\bibliography{references}

\end{document}